# Audio Recording Device Identification Based on Deep Learning

Simeng Qi, Zheng Huang
School of Electronic Information and Electrical Engineering
Shanghai Jiao Tong University
Shanghai, P.R.China
E-mail: {smart.qi, huang-zheng}@sjtu.edu.cn

Yan Li, Shaopei Shi
Institute of Forensic Science
Ministry of Justice
Shanghai, P.R.China
E-mail: {liyan, shisp}@ssfjd.cn

*Abstract*—**In this paper we present a research on identification of audio recording devices from background noise, thus providing a method for forensics. The audio signal is the sum of speech signal and noise signal. Usually, people pay more attention to speech signal, because it carries the information to deliver. So a great amount of researches have been dedicated to getting higher Signal-Noise-Ratio (SNR). There are many speech enhancement algorithms to improve the quality of the speech, which can be seen as reducing the noise. However, noises can be regarded as the intrinsic fingerprint traces of an audio recording device. These digital traces can be characterized and identified by new machine learning techniques. Therefore, in our research, we use the noise as the intrinsic features. As for the identification, multiple classifiers of deep learning methods are used and compared. The identification result shows that the method of getting feature vector from the noise of each device and identifying them with deep learning techniques is viable, and well-preformed.**

## I. INTRODUCTION

The identification of audio recording devices has numerous applications. In criminology and forensics, determining the audio recording device can help determining whether a certain record is from a proper device and thus determining its validity. In copyright disputes, finding out the actual ownership of a certain record may help deal with multiple claims of ownership. Also, if different pieces of one record show different recording devices, we can infer that the record may have been modified. However, at the same time, the advent of modern digital era adds up the difficulty and complexity to the identification. Thus, demand for efficient methods to assure the authenticity of audio signal is becoming more and more important.

Many people have done the identification of audio recording devices for numerous proposes in numerous conditions. Luca Cuccovillo et al. [1] used microphone classification to perform audio tampering detection, and the underlying algorithm was based on blind channel estimation and applied to detect a specific type of tampering. Constantine Kotropoulos et al. [2] performed research on mobile phone identification using recorded speech signals, and they used Mel frequency cepstral coefficients extracted from recorded speech signals to train a Gaussian Mixture Model with diagonal covariance matrices, thus providing templates for each device. Also, Ling Zou et al. [3] had similar ideas and utilized Gaussian mixture model-universal background model as the classifier, and showed that Mel frequency cepstral coefficients are more effective than Power-normalized cepstral coefficients. Xavier Valero et al. [4] compared Gammatone cepstral coefficients to Mel frequency cepstral coefficients in non-speech audio classification and found the GTCC more effective than MFCC, especially at low frequencies. Some people focus on the identification method themselves. Daniel Garcia-Romero et al. [5] proposed a method of automatic identification of acquisition devices when only get access to the output speech recordings, which used a support vector machine classifier to perform closed-set identification experiments and focused on two classes of acquisition devices. Robert Buchholz et al. [6] extracted a Fourier coefficient histogram of near-silence segments of the recording as the feature vector and used machine learning techniques for the classification. As for the features used to perform classification, people have different ideas. Yannis Panagakis et al. [7] chose random spectral features extracted from each speech signal as an intrinsic fingerprint for device identification, and Constantine Kotropoulos [8] chose the sketches of spectral features. Many other people chose the background noise of the audio recording devices to be the feature. Sohaib Ikram et al. [9] had a great idea about leakage signal, which is actually in the removed noise from speech enhancement, and we find the idea really inspiring. Huy Quan Vu et al. [10] identified microphone from noisy recordings by using representative instance One Class-Classification approach, and proposed a representative instance classification framework to improve performance of OCC algorithms. Chang-Bae Moon et al. [11] proposed an audio recorder identification method as one of digital forensic technologies, as well as a new feature reduction method, where Wiener filter was used to extract noise sounds of recorders and their features were extracted by MIRtoolbox. Rachit Aggarwal et al. [12] used features based on estimates of noise associated with recordings and classified them using sequential minimal optimization based Support Vector Machine. In this paper, we choose background noise as the feature, use classifiers of deep learning methods, improve former methods with new ideas and experiments, and show a pretty satisfying result.

Since in most cases, only the recorded audio signals are accessible, the identification should be based totally on the recorded audio signals themselves. This fact makes the problem pretty challenging since the audio signals we can get contain two parts: the speech signals and the noise signals, and the speech signals have their own variability based on the content. Usually, it's the speech which is regarded as the information to be passed that people mainly care about, and a great amount of researches have been dedicated to getting higher SNR. There are many speech enhancement algorithms to improve the quality of the speech signals, which can be seen as reducing the noise. However, noises can be regarded as the intrinsic fingerprint traces of an audio recording device. These digital traces can be characterized and identified by new machine learning techniques. Therefore, the noise can serve as the intrinsic features for the identification.

Deep learning is a set of algorithms in machine learning. It attempts to model high-level abstractions in data by using model architectures, which are composed of multiple non-linear transformations. Softmax regression is an important method in deep learning area for multi-class classification. The Softmax regression model generalizes logistic regression to classification problems where the class label can take on more than two possible values. This will be useful for problems where the goal is to distinguish between multiple outputs, in our case, multiple audio recording devices. Multilayer perceptron (MLP) is also appropriate for our situation. It is a modification of the standard linear perceptron and can distinguish data that are not linearly separable. With a feedforward artificial neural network model, MLP maps input data sets onto appropriate output set. It uses a backpropagation algorithm, and turns out to be a pretty proper algorithm for any supervised learning pattern recognition process.

Thus we present a research on identification of audio recording devices from background noise in the audio signals. Multiple classifiers of deep learning methods are used and compared to perform identifications. Furthermore, we also perform several enhanced methods such as model averaging and voting model to get better results.

The remaining paper is structured as follows: In Section II we introduce the audio files collection and the pre-processing of the dataset. Section III presents the methodology, including the background noise extraction, feature extraction, and classifiers. The whole experiment processing, as well as the results and corresponding analysis are shown in Section IV and we get the conclusion in Section V.

## II. DATASET

We use nine devices to record audio signals and the classification is among these nine categories. The devices are listed as Table I.

TABLE I. DEVICES

| Brand | Model |
|---|---|
| aigo | R5511 |
| Allbar | UB10 |
| HYUNDAI | HYV-B10 |
| JWD | DVR-601 |
| LG | g2 |
| OLYMPUS | WS-811 |
| PHILIPS | SA2SPK04K/93 |
| Shinco | V-21 |
| SONY | NWZ-B172F |

For each device, we generate three recordings and label them with 1, 2 and 3. Each recording's length varies from six minutes to seven minutes. For each audio file, we randomly split it and generate segments to be our dataset, with 4096 sample points in each miner segment. Among the three recordings for each equipment, the first two are used to generate the training set and each recording results in 1000 segments. The third file is used for the testing set, and each one generates 100 segments. The data instances extracted from the recordings are assigned with the label of the device by the file name. In this way, a total of 18000 identification trials for training and 900 for testing are obtained. Each of the segments will form data instances with feature vector.

## III. BACKGROUND NOISE CLASSIFICATION METHODOLOGY

The overview of our experiment is shown in Fig. 1.

### A. Background Noise Extraction

As long as we have the audio segments, we need to extract the intrinsic characteristic of the device, which in our case, is the noise of the signals. The input signal is a passively received audio signal which is the sum of speech signal and noise signal. This can be expressed as:

$$s_n = f_n + e_n \qquad (1)$$

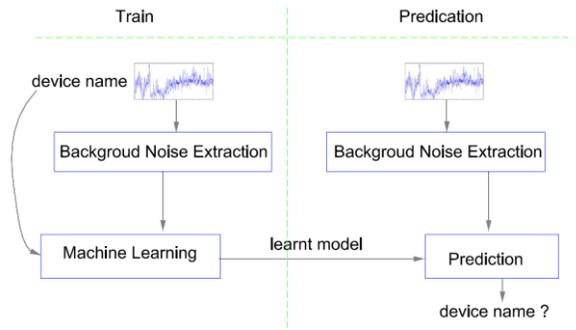

Figure 1. Overview

in which, time $n$ is equally spaced. $s_n$ is the audio signal we get from former processing. $f_n$ is the speech signal, which is the information people usually care about. $e_n$ is the noise signal that we need here to identify the device.

A great amount of researches have been dedicated to getting higher SNR. So there are many speech enhancement algorithms to improve the quality of speech signals, which can be seen as reducing the noise. In our experiment, we get the de-noised version of the input signal, which is expressed in the equation as $f_n$.

We use an automatic de-noising process of a 1-D signal using wavelets. The de-noise objective is to suppress the noise part of the signal $e$ and to recover $f$.

After doing the subtraction, we can obtain the noise signal $e_n$.

For the baseline, which is to generate features directly from the original audio signals, this part is unnecessary. We just assign $s_n$ to $e_n$.

### B. Feature Exaction

After obtaining the signal $e_n$, we need to turn them into feature vectors for the classification. Our feature extraction is based only on frequency domain features of a recording, which are mathematical representation reflecting characteristics of audio signal. To do so, we extract a Fourier coefficient histogram of the signal as the feature vector, which has powerful descriptive capability for audio signals. The corresponding Fourier coefficients for all those segments are summed up to yield a Fourier coefficient histogram that is then used as the global feature vector.

This process can be expressed as:

$$F(e_n) = FFT(e_n) \qquad (2)$$

After the Fast Fourier Transform, we normalize it as:

$$N(F(e_n)) = log(F(e_n) + 1) \qquad (3)$$

In this way, the segments of the speech recording are represented by a point in a high-dimensional vector space. The data after the transformation are saved as the dataset **x**. For training, **x** is a matrix of 2049*18000 and for testing, **x** is a matrix of 2049*900. These matrixes provide the features that we need. At the meantime, the file names of the audios are written into vector **y** and recorded in a map, which are also assigned with the label of the device by the file name.

### C. Classifiers

Deep learning is a set of algorithms in machine learning. It can model high-level abstractions in data by using model architectures, which are composed of multiple non-linear transformations. An observation, in our case, audio signal, can be represented as a vector, and thus can be processed by multiple standard algorithms.

We'll give a brief introduction to the methods we used in our experiments. These are not necessarily independent. To obtain a better solution, some mixed-up may get surprising result.

*1) Softmax:* Softmax regression model is one of the models in the field of deep learning, which generalizes logistic regression and is expanded to perform classification among more than two classes.

For a training set $\{(x^{(1)}, y^{(1)}),...,(x^{(m)}, y^{(m)})\}$, where the input feature $x_i \in R_{n+1}$. The difference from the logistic regression here is that instead of $\{0, 1\}$, we now have $y_i\{1, 2,...,k\}$ so that the label $y$ can be set among $k$ different value.

For a given test input **x**, we need our function to estimate the probability that $p(y=j/x)$ for $j = 1...k$, and thus to determine which label should $y/x$ be. So, our hypothesis is to output a $k$-dimensional vector, each element in the vector presents the estimated probability for one of the $k$ possible values of $y_i$, and these $k$ probabilities should sum to 1. The hypothesis $h_\theta$ can be described as:

$$h_\theta(x^{(i)}) = \begin{bmatrix} p(y^{(i)} = 1 \mid x^{(i)}; \theta) \\ p(y^{(i)} = 2 \mid x^{(i)}; \theta) \\ ... \\ p(y^{(i)} = k \mid x^{(i)}; \theta) \end{bmatrix} \qquad (4)$$

The cost function is:

$$J(\theta) = -\frac{1}{m}[\sum_{i=1}^{m}\sum_{j=1}^{k} 1\{y^{(i)} = j\} \log \frac{e^{\theta_j^T x^{(i)}}}{\sum_{l=1}^{k} e^{\theta_l^T x^{(i)}}}] + \frac{\lambda}{2}\sum_{i=1}^{k}\sum_{j=0}^{n} \theta_{ij}^2 \qquad (5)$$

in which $1\{*\}$ is called indicator and can be calculated as $1\{a\ true\ statement\} = 1$ and $1\{a\ false\ statement\} = 0$. In the cost function, the former part is the naive cost and the later part is a weight decay term which is to penalize large values of the parameters. To get a working implementation of Softmax regression, we will minimize $J_\theta$.

*2) MLP:* Multilayer perceptron (MLP) is another model in the field of deep learning. It maps input data sets onto output data sets. A MLP has multiple layers, where nodes connecting to every node in the next layer and the last layer is the output we need. With these connection, MLP has the ability to classify data which are not linearly separated.

Each node in MLP, except for those in the input layer, has a nonlinear activation function that maps the weighted inputs to the outputs. Apart from the input and output layers, MLP consists of one or more hidden layers. Since the activation functions are not linear, these layers cannot be reduced to the standard two-layer input-output model. In this way, it is considered a deep neural network.

Each node in one layer connects to every node in the following layer, with a certain weight. These weight will be changed after each piece of data is processed, by minimizing the error, which is presented as:

$$\varepsilon(n) = \frac{1}{2}\sum_i (d_i(n) - y_i(n))^2 \qquad (6)$$

where $d_i$ is the target value and $y_i$ is the produced value. In this way, MLP utilizes a supervised learning technique called backpropagation.

*3) CNN:* In machine learning, a convolutional neural network (CNN) is a type of feed-forward artificial neural network. Convolutional networks were inspired by biological processes and are variations of MLP designed to use minimal amounts of preprocessing.

CNN consists of multiple layers of small neuron collections. These collections process portions of the input data, and the outputs of these collections are then tiled so that they overlap, to obtain a better representation of the original data. This is repeated for every such layer.

Compared to other classification algorithms, CNNs use relatively little pre-processing. This means that the network is responsible for learning the filters that in traditional algorithms were hand-engineered. The lack of dependence on prior knowledge and human effort in designing features is a major advantage for CNNs. Considering this superb feature of CNN, we also performed several experiments without the noise-extraction using wavelets.

*4) Model Averaging:* Since as a baseline, we put our generated data into several hidden layers all together, and these data are actually not consecutive, it is reasonable to consider putting them separately through separate hidden layers, and then concatenating them all together for the succeeding processes.

*5) Voting Model:* One miner signal segment is easily to get biased randomly. So we consider to generate multiple sets of segments as input data for our classification. After all the predictions, we sum the results up and let them vote for the final classification result.

## IV. EXPERIMENT AND RESULT

We did several experiments to compare classifiers, and to compare parameters in one certain classifier. Open source code, with which the most important results and figures can be reproduced, is available at https://github.com/SMartQi/identification.

Table II shows the comparison of Softmax classifier and MLP classifier, according to the processes shown in Fig. 2. This indicates that MLP beats Softmax classifier in audio device classification in this certain condition, and that three hidden layers seem to work better, but not necessarily more layers leads to better result.

Table III shows the results without performing background noise extraction first before classification, according to the processes shown in Fig. 3. No matter which classifier is used, background noise signal shows better accuracy than the recording signal itself. Also, both Softmax and MLP have certain ability to extract features from the original recording signals automatically, and perform the classification all by themselves. As we introduced, compared to other classification algorithms, CNNs use relatively little pre-processing. This means that CNN network is responsible for learning the filters that in traditional algorithms were hand-engineered, and does not care so much about whether the input data have noise mixed. Notice, the **noise** here is not the noise signals in the audio signals. On the contrary, features should be generated from our noise signals, and the **noise** here refers to the speech signal, which is not helpful to the classification.

TABLE II. CLASSIFIERS COMPARISON

| Classifier | Hidden Layer Number | Accuracy (%) |
|---|---|---|
| Softmax | - | 87 |
| MLP | 1 | 92 |
| MLP | 2 | 91 |
| MLP | 3 | **93** |

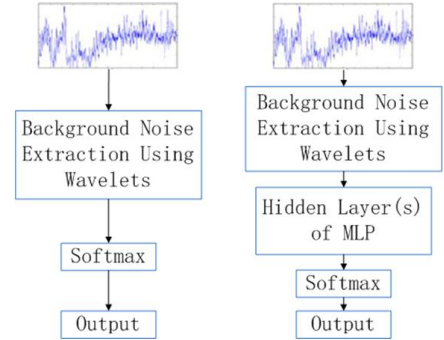

Figure 2. Processes of Softmax and MLP

TABLE III. RESULT OF NO BACKGROUND NOISE EXTRACTION

| Classifier | Accuracy (%) |
|---|---|
| Softmax | 79 |
| MLP | 84 |
| CNN | **90** |

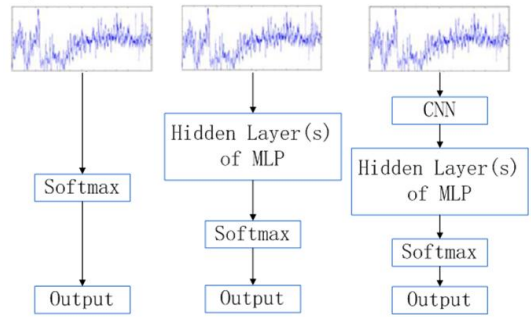

Figure 3. Processes without Background Noise Extraction

Table IV shows the comparison of whether to introduce model averaging, according to the process shown in Fig. 4. Unfortunately, the model averaging strategy does not have significant improvement.

Table V shows the comparison of whether to introduce the voting model, according to the process shown in Fig. 5. In these experiments, we set three hidden layers. We can clearly see that more voters lead to higher accuracy. But notice that more voters also lead to more time to generate datasets and perform classification.

TABLE IV. MODEL AVERAGING COMPARISON

| Hidden Layer Number | Model Averaging | Accuracy (%) |
|---|---|---|
| 1 | yes | 92 |
| 1 | no | 92 |
| 2 | yes | **92** |
| 2 | no | **91** |
| 3 | yes | 93 |
| 3 | no | 93 |

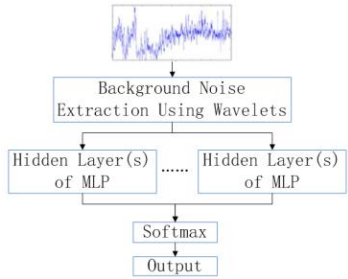

Figure 4. Model Averaging

TABLE V. VOTING MODEL COMPARISON -

| Voter Number | Accuracy (%) |
|---|---|
| 3 | 96 |
| 4 | 98 |
| 5 | 99 |

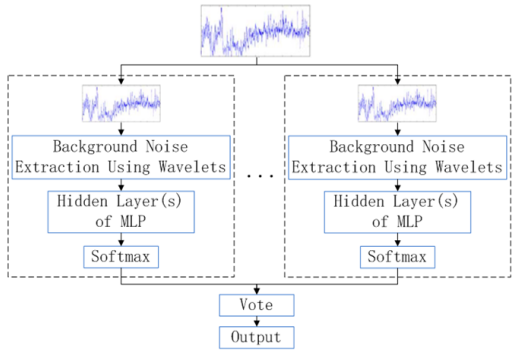

Figure 5. Voting Model

Table VI shows the detailed result of MLP classification. The confusion matrix result is given as Fig. 6. Fig. 7 is the dimensionality-reduced result using t-distributed Stochastic Neighbor Embedding (t-SNE). We can clearly see that the classification result is pretty satisfying. One thing to notice is that the error is not evenly distributed among all devices. Device aigo and PHILIPS get pretty high recall, as well as good precision and f1-score. These are also shown in Fig. 6 with dark blocks and in Fig. 7 with clustered dots. It's relatively easy to confuse device Allbar and Shinco, as shown in Fig. 6 that the yellow block is relatively obvious for their two corresponding positions. Some mixed-up dots for them are shown in Fig. 7.

TABLE VI. MLP TEST RESULT

| Device | Precision | Recall | F1-score |
|---|---|---|---|
| 1 | 0.96 | 0.99 | 0.98 |
| 2 | 0.83 | 0.95 | 0.89 |
| 3 | 0.98 | 0.92 | 0.95 |
| 4 | 0.85 | 0.95 | 0.90 |
| 5 | 0.99 | 0.89 | 0.94 |
| 6 | 0.98 | 0.82 | 0.89 |
| 7 | 0.98 | 1.00 | 0.99 |
| 8 | 0.87 | 0.81 | 0.84 |
| 9 | 0.93 | 1.00 | 0.96 |
| average/total | 0.93 | 0.93 | 0.93 |

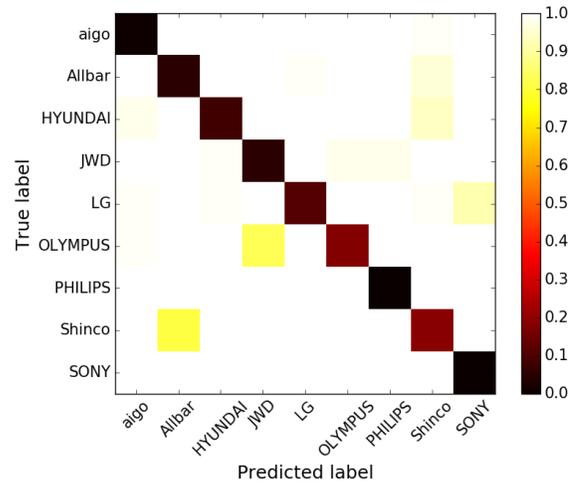

Figure 6. Normalized Confusion Matrix

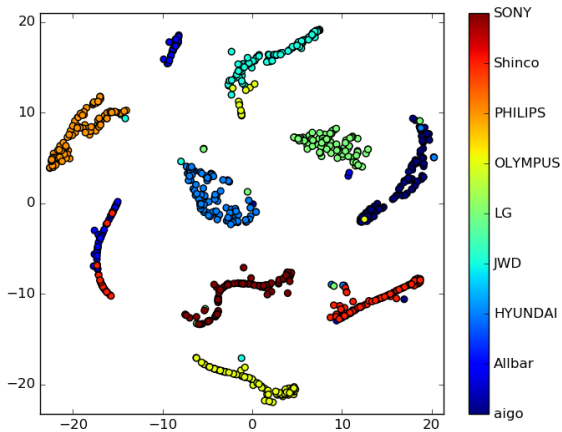

Figure 7. Dimensionality Reduction Presentation

## V. CONCLUSION

A research on identification of audio recording devices from background noise in the audio signals is presented. It showed the method of getting feature vectors from the noise of each device and identifying those using deep learning methods is viable, and well-preformed.

Although pretty much work has been done, many issues in the field of identification of audio recording devices are still open. Some other topics are still promising. On the one hand, our underlying theory is that the input signal is a passively received audio signal which is the sum of speech signal and noise signal, and by doing the subtraction, we can get the noise signal; on the other hand, the current de-noise methods pay more attention to the quality of the speech, instead of the noise signal. Thus, there may be a bias between the real noise signal and the signal after the subtraction. More work can be done in the direction of getting purer noise signal.

## ACKNOWLEDGMENT

This work has been supported by the "12th Five-Year Plan" National Science and Technology Support Program under the grant 2012BAK16B05. The authors would like to thank our co-worker Jingyao Luo, who helped a lot during the collection of audio recording materials.

## AUTHORS' BACKGROUND

| Your Name | Title* | Research Field | Personal website |
|---|---|---|---|
| Simeng Qi | master student | Machine Learning | |
| Zheng Huang | assistant professor | Machine Learning | http://cis.sjtu.edu.cn/index.php/Zheng_Huang |